\magnification=1200
\medskip
\centerline {\bf RESONANCE NLS SOLITONS AS BLACK HOLES} 
\medskip
\centerline {\bf  IN MADELUNG FLUID}
\medskip
\vskip 1.2cm
\centerline{O. K. PASHAEV$^{1,2}$\footnote\dag{e-mail: pashaev@vxjinr.jinr.ru;
pashaev@math.sinica.edu.tw} and JYH-HAO LEE$^{2}$\footnote{*}{e-mail:
leejh@ccvax.sinica.edu.tw}}
\medskip
\centerline{\it $^{1}$Joint
Institute for Nuclear Research, Dubna (Moscow), 141980, Russia}
\medskip
\centerline{\it $^{2}$Institute of Mathematics, Academia Sinica,
 Taipei 11529, Taiwan}

\bigskip

\centerline{Abstract}
\medskip
\par{{\sl  
A new resonance version of NLS equation is found and  
embedded to the
reaction-diffusion system,
equivalent to the anti-de Sitter valued
Heisenberg model, realizing a particular gauge fixing condition
of the Jackiw-Teitelboim gravity. The space-time points where dispersion 
change the sign correspond to the event horizon, and the soliton
solutions to the AdS black holes. The soliton with velocity
bounded above 
describes evolution on the hyperboloid
with nontrivial winding number   
and create under collisions the resonance states  with a
specific life time. }

\par \bigskip\bigskip
\par\noindent
{\bf 1. Resonance NLS equation.}
\bigskip\par
The connection between black hole physics and the theory 
of supersonic acoustic flow was established by Unruh$^1$ and 
has been developed to investigate the Hawking radiation and
other phenomena for understanding quantum gravity$^2$.   
Recently, the similar idea for simulation of quantum effects related to
 event horizons
and ergoregions but in superfluids, which in contrast to the usual liquids
 allow nondissipative motion of the flow, 
 was proposed$^3$. In this 
case a ''superluminally" moving inhomogeneity of the order
parameter like solitons,  provides black holes like quasi-equilibrium
states, exhibiting an event horizon. Although
Jacobson and Volovik considered  a simplified profile of 
soliton and mentioned unimportance of exact structure of the solution,
an  
exactly soluble model of solitons related to  black holes allows one
to describe the scattering process of black holes and corresponding 
quantum phenomena. Very recently, some attempt was done to describe
solitons of integrable models like the Liouville$^4$, the Sine-Gordon
$^{5,15}$ and the  Reaction-Diffusion system$^{6,7}$, as black holes of 
Jackiw-Teitelboim (JT) gravity. In the last paper the scattering of 
two identical soliton-like structures called $dissipatons$,  
relating to the black holes and creating  
the metastable state was considered. 
Continuing in this direction, in the present paper
we present a new integrable version of the Nonlinear 
Schr\"odinger equation (NLS) in 1+1 dimensions, admitting solitons 
with a rich resonance scattering phenomenology
and interpreted 
as black holes of JT gravity. 
\par
The NLS,  well known also as the Gross-Pitaevskiy 
equation, appears in the phenomenological description of superfluidity
of an almost ideal Bose gas$^8$.
In this case,  the squared modulus of the wave function $\bar \psi\psi$
is interpreted as the particle number density in
the condensate state, while the gradient of the phase is proportional
 to the 
superfluid velocity $v = \nabla \arg \psi$. The similar
hydrodynamical interpretation  for the decomposition 
of quantum mechanical wave function was considered 
by Madelung
long time ago$^9$. 
It can be extended to the NLS, representing a generic envelope equation 
and appearing in the wide range of 
phenomena, with  many applications. Besides superfluidity,
 the most popular on is probably in the nonlinear optics, describing
light propagation in optical fibers$^9$. 
The decomposition shows that quadratic dispersion term in the NLS results
from contribution of two terms: the superflow (classical drift) dispersion 
and the 
so called ''quantum potential"
$$U_{Q}(x) = 2 |\psi|_{xx}/ |\psi|.\eqno(1.1)$$
The potential was introduced by de Broglie$^{10}$ and explored by 
D. Bohm$^{11}$
to make a hidden-variable
theory and is responsible for
producing the quantum behavior, so that all quantum features
are related to its special properties. 
In more recent context it is considered in the stochastic mechanics 
as producing non-classical diffusion$^{12}$. 
Relation of non-classical motion with the {\it internal} spin motion or 
the {\it zitterbewegung} 
was established in a series of papers$^{13}$.
 In the optical context 
the superflow term corresponds to the geometrical optics, while 
the quantum potential to the diffraction,
 which leads to the spontaneous pattern formation and allows the
quantum effects even at room temperature and macroscopic scale.
We emphasize that 
  existence of the envelope solitons 
for nonlinear equations is
possible
only with this potential, since its dispersive contribution is 
exactly compensated by
the nonlinearity.
Changing the sign of  quantum potential contribution to the NLS,
modify the behavior of 
solitons drastically. 
The potential $U_{Q}$ is invariant under 
the complex constant rescaling transformations,
$\psi(x,t) \rightarrow \lambda \psi(x,t), \, \lambda \in {\cal C}$,
and thus does not depend on the strength of
the wave, associated with soliton , but depends only of the 
form of the wave. Therefore, its effect could be large even for the
well separated and far enough solitons.
\par
Below we consider  the NLS soliton propagation in
 the quantum potential (1.1):
$$
i\partial _{t}\psi  +  \partial
^{2}_{x}\psi +  {\Lambda \over 4}|\psi|^{2}\psi  = 
2 {|\psi|_{xx} \over |\psi|}\psi, \eqno(1.2) $$
which we call the {\it resonance Nonlinear Schr\"odinger equation}
(RNLS). 
It could corresponds to
the response of a hypothetical resonance medium to an action of a
quasimonochromatic wave with complex amplitude $\psi(x,t)$,
 which is slowly 
varying function of the coordinate and the time.
Eq.(1.2) can be considered as
the third integrable version of NLS, intermediating 
 between the defocusing and 
focusing cases (repulsive and attractive non-ideal Bose gas correspondingly).  
The additional term in the
right hand side of NLS (1.2), depending on the form of the amplitude profile
of the wave, can be interpreted also as a specific electrostriction pressure
or the diffraction ''force"$^{14}$. Worth to note that the model (1.2) is 
integrable
only  with the chosen coefficient 2 on the r.h.s, but the resonance
properties can be valid as well in a more general case.

\bigskip\par\noindent
{\bf 2. Madelung fluid and the reaction-diffusion system.}
\bigskip\par
By decomposing
the wave function $\psi \in {\cal C}$ , $\psi = \exp (R - iS)$, $\bar\psi =
\exp (R + iS)$, in terms of two real functions $R, S \in {\cal R}$, the model 
(1.2) can be represented as the Madelung fluid$^9$
$$-\partial_{t} R + \partial^{2}_{x} S + 
2 \,\partial_{x} R \,\,\partial_{x} S = 0,\eqno(2.1a) $$
$$
-\partial_{t} S + \partial^{2}_{x} R + (\partial_{x}R)^{2}
+ (\partial_{x} S)^{2} - {\Lambda \over 4}e^{2R} = 0. \eqno(2.1b)$$
Introducing two new real functions
$e^{+} = \exp (R + S), \, -e^{-} = \exp (R - S)$, 
or
$$ -e^{+}e^{-} = e^{2R} =
 |\psi|^{2},\,\,\, S = {1 \over 2} \ln {e^{+} \over -e^{-}} = {1 \over 2i} 
\ln {\bar\psi \over \psi},\eqno(2.2)$$
we have the system
$$
-\partial _{t}e^{+}  +  \partial
^{2}_{x}e^{+} + {\Lambda \over 4}e^{+}e^{-}e^{+}  = 0, \eqno(2.3a) $$
$$
+\partial _{t}e^{-}  +  \partial
^{2}_{x}e^{-} + {\Lambda \over 4}e^{+}e^{-}e^{-}  = 0, \eqno(2.3b) $$
representing a particular form of a 2-component
 reaction - diffusion (RD) system. We note that 
 unusual  negative value for 
 diffusion coefficient in the second equation is crucial for the
existence of Hamiltonian structure and the
integrability of
the model. In this case the system (2.3) is
time reversible $t \rightarrow -t$ and
invariant under  the   global $SO(1, 1)$
transformations  $e^{\pm}\rightarrow e^{\pm
\alpha} e^{\pm}$. It admits
solution with exponentially growing and decaying 
components,
$$ e^{\pm} =
 \pm({8 \over -\Lambda})^{1 \over 2}k e^{\pm [({1 \over 4}v^{2} + k^{2})t - 
{1 \over 2}vx]} 
\cosh^{-1} [k(x - v t - x_{0})], \eqno(2.4)$$
 but with perfect solitonic shape for the $O(1,1)$ scalar
product 
$$ -e^{+}e^{-} = |\psi|^{2} = 
{8 \over -\Lambda}k^{2} \cosh^{-2} [k(x - v t - x_{0})].\eqno(2.5)$$
By analogy with the dissipative structures in the pattern
formation we called these dissipative soliton solutions  as $dissipatons$$^6$.
\par 
Using (2.2) we can find one-soliton solution of Eq.(1.2) corresponding to
the one-dissipaton solution (2.4)
$$ \psi = 
({8 \over -\Lambda})^{1 \over 2}k 
{e^{-i[({1 \over 4 }v^{2} + k^{2})t - {1 \over 2}vx]} 
\over \cosh k (x - vt)}.\eqno(2.6)$$
As we  see  the role of the ''quantum potential"
 is to change the dispersion for the NLS. Indeed, when
$U_{Q} = 0$, Eq.(1.2) is the so-called defocusing NLS, admitting the 
``dark" soliton solution with non-vanishing boundary values. 
While with $U_{Q} \ne 0$, 
we have the ``bright" soliton (2.6),
with vanishing boundaries, which usually appears in 
the focusing NLS.
\bigskip\par\noindent
{\bf 3. The gravitational interpretation.}
\bigskip\par
Dissipatons are related to the black hole solutions of the Jackiw-Teitelboim
gravity$^{4,7}$  and provide interesting tools to study nonperturbative
sector of the General Relativity. 
Defining the 2-dimensional metric tensor in terms of 
Einstein-Cartan zweibeins
$$g_{\mu\nu} = e^{a}_{\mu}e^{b}_{\nu} \eta_{ab} = 
{1 \over 2}(e^{+}_{\mu}e^{-}_{\nu}
+ e^{+}_{\nu}e^{-}_{\mu}),\eqno(3.1)$$ 
where $e^{\pm}_{\mu} = e^{0}_{\mu} \pm e^{1}_{\mu} = 
(e^{\pm}_{0}, e^{\pm}_{1})$, $\eta_{ab} = diag(-1,1)$, 
and
$$ e^{\pm}_{0} = \pm{\partial \over \partial x} e^{\pm},\,\,\, 
e^{\pm}_{1} \equiv e^{\pm}, \eqno(3.2)$$
such that
$$g_{00} = - {\partial e^{+} \over \partial x}
{\partial e^{-} \over \partial x}, \,\,\, 
g_{11} = e^{+}e^{-},\,\,\,g_{01} = 
{1 \over 2}({\partial e^{+} \over \partial x}e^{-}
- e^{+}{\partial e^{-} \over \partial x}),\eqno(3.3)$$ 
we find that for $e^{\pm}$ satisfying (2.3)
the metric describes  
 two dimensional pseudo - Riemannian  
space-time
with  constant curvature $\Lambda$:
$$R = g^{\mu\nu} R_{\mu\nu} = \Lambda.\eqno(3.4)$$
This, low dimensional (lineal) gravity model
 is known as the Jackiw-Teitelboim
model$^4$.  It can be used also to describe the 
S-wave sector of the extremal D = 4 supersymmetric black hole
 solutions of models
with specific dilaton coupling$^{15}$.
When the curvature vanishes, $\Lambda = 0$, the nonlinear 
term in Eqs.(1.2) and (2.3) disappears. 
Then the system (2.3) becomes the decoupled 
linear heat equation and the time-reversal one,
 while (1.2) reduces to the 
unusual, sign indefinite dispersive modification of 
the linear Schr\"odinger equation:
$$
i\partial _{t}\psi  +  \partial
^{2}_{x}\psi - 
2 {|\psi|_{xx} \over |\psi|}\psi = 0. \eqno(3.5) $$
 Worth to note that the classical theories
corresponding to the usual quantum mechanical
 Schr\"odinger equation and to our
modified model (3.5) would be equivalent, since the ''quantum potential"
is proportional to $\hbar$, 
and  in the $\hbar \rightarrow 0$ limit
both models lead to the same Hamilton-Jacobi equations. 
It turns out that this model 
is relevant to the black hole solutions of the 
CGHS string-inspired gravitational theory$^{16}$.
\par
The gauge fixing condition (3.2) is equivalent to 
the classical $SO(2,1)/SO(1,1)$ Heisenberg model on the 
anti-de Sitter space ($\Lambda < 0$),
$$\partial
_{0}{\bf s} = {\bf s
\wedge
    \partial }^{2}_{x}{\bf s}, \eqno(3.6)$$
where $e^{\pm}_{\mu}$ are local coordinates in the tangent plane
$\partial_{\mu} {\bf s} = ({-\Lambda \over 8 })^{1 \over 2}
(e^{+}_{\mu} {\bf n}_{-} - e^{-}_{\mu}{\bf n}_{+})$,
so that the metric tensor is
$$g_{\mu \nu} = ({2 \over -\Lambda})
(\partial_{\mu}{\bf s}\partial_{\nu}{\bf s}),\eqno(3.7)$$
and ${\bf s}^{2} = - (S^{1})^{2} + (S^{2})^{2} - (S^{3})^{3} = -1$.
Actually in Sec.7 we show that  (3.6) is gauge equivalent to (2.3).
\par
Formulas (3.3) allow us to establish a correspondence between the geometrical 
and physical 
characteristics of the model. Indeed, in terms of $\psi$ variables 
the metric tensor is given by
$$g_{00} = 2({\partial |\psi| \over \partial x})^{2} 
- {\partial \bar\psi \over \partial x} {\partial \psi \over \partial x}
,\,\,g_{11} = - |\psi|^{2},\,\,
g_{01} = {i \over 2}({\partial \bar\psi \over \partial x }\psi -
\bar\psi {\partial \psi \over \partial x}),\eqno(3.8)$$
so that $g_{00}$ component is the dispersive part
of energy density, while $g_{11}$ and $g_{01}$ the mass and momentum
densities correspondingly. The mass, momentum and energy conserved quantities
$$ M = - \int^{\infty}_{-\infty} e^{+}e^{-} \, dx =
          \int^{\infty}_{-\infty} |\psi|^{2} \, dx, \eqno(3.9a)$$

$$ P = - \int^{\infty}_{-\infty} (e^{+}\partial_{x}e^{-} -
\partial_{x}e^{+} e^{-}) \, dx =
 i \int^{\infty}_{-\infty} ( \partial_{x}\bar\psi \,\psi - \bar\psi
\partial_{x}\psi ) \, dx, \eqno(3.9b)$$

$$ E = 2\int^{\infty}_{-\infty} [\partial_{x}e^{+}\partial_{x}e^{-} 
-{\Lambda \over 8}(e^{+} e^{-})^{2}] \, dx =
 2\int^{\infty}_{-\infty} [\partial_{x}\bar\psi \partial_{x}\psi -
 2 \partial_{x} |\psi|\partial_{x} |\psi| 
- {\Lambda \over 8} |\psi|^{4}]
 \, dx, \eqno(3.9c)$$
for one-dissipaton solution (2.4) or the one-soliton solution (2.6)
 are
$$M = {16 \over -\Lambda}|k|, \,\,\,\,\, P = Mv, \,\,\,\,\,
 E = {Mv^{2} \over 2} + {\Lambda^{2} \over 384}M^{3}.\eqno(3.10)$$
These expressions show that it
can be interpreted as non-relativistic quasi-particle of non-negative
 mass M and momentum P, with the positive rest energy 
$ E_{0} = E( v = 0) = {\Lambda^{2} \over 384} M^{3}$.
\bigskip\par\noindent
{\bf 4. Resonance dispersion and black holes.} 
\bigskip\par
As we will see below
an interaction of these particles  
leads to the creation and 
annihilation processes, forming resonance states. 
The existence of these 
states  follows from the form of dispersion part of energy density (3.9c),
written in terms of variables (2.2) 
$$\epsilon_{0} \equiv 2(\partial_{x}\bar\psi \partial_{x}\psi -
 2 \partial_{x} |\psi|\partial_{x} |\psi|) = 2[(\partial_{x}S)^{2} - 
(\partial_{x}R)^{2}]
e^{2R}.\eqno(4.1)$$
For the NLS case, when $U_{Q} = 0$, 
we have the positive definite dispersion energy
$$\epsilon_{0} \equiv 2\partial_{x}\bar\psi \partial_{x}\psi  = 
2[(\partial_{x}S)^{2} + (\partial_{x}R)^{2}]
e^{2R},\eqno(4.2)$$
while in contrast
for $U_{Q} \ne 0$, from (4.1) it becomes indefinite. Function
$\epsilon_{0}$ changes the sign at space-time points where
$$(\partial_{x}S)^{2} - (\partial_{x}R)^{2} = (\partial_{x}S - \partial_{x}R)
(\partial_{x}S  + \partial_{x}R) = 0,\eqno(4.3)$$
or 
$\partial_{x}S = \pm \partial_{x}R$.
Now, if we compare (4.1) with (3.8), so that the dispersion energy density  
has geometrical meaning of the metric tensor
component $\epsilon_{0} = -2g_{00}$, then conditions (4.3) are equivalent
to  existence of the event horizon space-time points $(x_{H}, t_{H})$, where 
$g_{00}$, change the sign.
 In this way we relate the resonance dispersion 
 of NLS with the ''quantum potential"  (1.2), to the 
event horizon in two dimensional space-time. 
The existence of  event horizon
indicates the nontrivial causal structure of the space-time 
and the black hole
phenomena. In fact, if we calculate the metric (3.8)
 for one-soliton solution (2.6)
$$ds^{2} = {8 \over -\Lambda}[(k^{2}\tanh ^{2} k(x - vt) -
 {1 \over 4}v^{2})(dt)^{2} - (dx)^{2} - v dx dt]|\psi|^{2},\eqno(4.4)$$
it shows a horizon singularity at
$$\tanh k(x - vt) = \pm {v \over 2k},\, \,\,\,only\,\,\,\, if,\,\,\,\, 
|v| \le 2|k| \equiv |v_{max}|.\eqno(4.5)$$
Consequently, a black hole dissipaton cannot move faster
 than the maximal value of the 
velocity $|v_{max}| = 2|k|$. 
In this case the metric can be transformed to the Schwarzschild type
form  and shows the causal structure
 in terms of Kruskal-Szekeres coordinates$^7$. 
\bigskip\par\noindent
{\bf 5. Hydrodynamical interpretation.}
\bigskip\par 
The Madelung fluid representation gives 
simple hydrodynamical explanation of
 of the resonance states and the event horizon existence. 
In (2.2) representation we introduce the fluid density 
$\rho \equiv e^{2R} = |\psi|^{2}$
and the local velocity
$ V \equiv -2\partial_{x}S$,
describing the center of mass motion.
Then,
characterizing the internal motion referred to the center of mass frame,
the stochastic diffusion or 
$the\,\, zitterbewegung$$^{13}$ 
 non-classical contribution from the ''quantum potential",
 can be described by 
velocity $V_{Q} \equiv {\partial_{x} \rho \over \rho}$.
For the normal Madelung fluid (the NLS, when $U_{Q} = 0$)
 the dispersion energy
density (4.2) is just the sum of kinetic energies of
these two motions  
$$\epsilon_{0} = ({\rho V^{2} \over 2} + {\rho V^{2}_{Q} \over 2}).\eqno(5.1)$$
In contrast, for the non-vanishing $U_{Q} \ne 0$, the density  (4.1)
 is the difference
$$\epsilon_{0} = ({\rho V^{2} \over 2} - {\rho V^{2}_{Q} \over 2}).
\eqno(5.2)$$
In this hydrodynamical representation, 
the metric tensor (3.8)
has the form
$$g_{00} = {1 \over 4}\rho (V^{2}_{Q}  - V^{2}),\,\,
g_{11} = - \rho, \,\, g_{01} =  {1 \over 2} \rho V,\eqno(5.3)$$
which 
is similar to the ADM split of a (1+1)-dimensional
Lorentzian spacetime corresponding to the so-called acoustic metric, 
derived by Unruh$^1$ for the sound waves in a fluid$^{2}$.
Then, the existence of resonance states or the event
 horizon has the meaning of 
the equality of the center of mass and internal motion velocities
$V = \pm V_{Q}$.
For the one-soliton solution (2.6) velocity of the classical center of
mass motion  is constant and coincides with the soliton propagation velocity 
$V = v$,
while the ''quantum" velocity is
$V_{Q} = -2k \tanh k(x - vt)$.
The last one is bounden above by the constant value 
$|V_{Q}| \le 2|k|$,
why  the velocities compensation 
is possible only in this region. Then, the compensation condition
$V = \pm V_{Q}$ is equivalent to the event horizon Eq. (4.5).
\par
To derive the black hole  
we first rewrite the metric (5.3) in the moving frame  
$(\xi,t) = (x - vt, t)$ with constant velocity $v$.
Then it  has convenient form 
in terms of the new ''shifted" local velocity 
$W(\xi,t) = 2v - V(\xi,t)$:
$$\tilde g_{00} = {1 \over 4}\rho (V^{2}_{Q}  - W^{2})
,\,\,
\tilde g_{11} = - \rho, \,\, \tilde g_{01} =  -{1 \over 2} \rho W.
\eqno(5.4)$$.
\par For a one-soliton solution (2.6) or dissipaton (2.4),
moving with a constant velocity $v$, 
we have $W = v$ and the metric (5.3), where $V = v = const.$,
but with stationary space-time geometry
due to the time independence of
$\rho = \rho(\xi), \,\,\, V_{Q} = V_{Q}(\xi)$. 
\par
In the general case the metric contains the off-diagonal terms.
The time synchronization of this space-time is possible if function 
$2W/(V_{Q}^{2} - W^{2})$ is integrable.
 Then we can define new time coordinate
$d\tau = dt - 2W/( V^{2}_{Q} - W^{2})d\xi $ 
and get the metric
$$ds^{2} = \rho [ {1 \over 4}(V^{2}_{Q}  - W^{2}) (d\tau)^{2}
 - {V^{2}_{Q} \over V^{2}_{Q} - W^{2}}(d\xi)^{2}]. \eqno(5.5)$$
Following the same arguments as for the black holes,
 the Hawking temperature
can be derived from this metric.
 In particular case of one soliton solution
everything can be done explicitly. Synchronization of the stationary metric 
is given by the above 
 time transformation integrated as 
$$ \tau = f(\xi,t) =
 t + {v \over 2k^{3}(1 - \gamma^{2})}\left[- k\xi + {1 \over 2|\gamma|}
\ln \left|{|\gamma| +
 \tanh k\xi \over |\gamma| - \tanh k\xi}\right|\right],$$
where $|\gamma| \equiv |v/2k| < 1$. 
The Hawking temperature in this case 
is
$T_{H} = {1 \over 2\pi} k^{2}(1 - \gamma^{2})$.
\par
Comparing (5.2) with (3.9c) and (3.10) we can see that the kinetic part of the
internal motion gives the negative sign 
contribution to the rest energy $E_{0}$, 
but due to the positive potential part, the resulting energy is positive.
This is the reason why we can
have the resonance behavior for our model, but not for the NLS. Indeed,  
decay of a soliton in the rest means creation of a pair
with the positive energy, which is allowed only if the rest energy is positive. 
From conservation of the mass (3.9a) follows that the mass defect for soliton
decay is always zero,
$\Delta M = M  - (M_{1} + M_{2}) = 0$. 
Then, the rest energy satisfies the condition
$$E_{0}= {\Lambda^{2} \over 384}M^{3} = 
{\Lambda^{2} \over 384}(M_{1} + M_{2})^{3}>  
 {\Lambda^{2} \over 384}(M^{3}_{1} + M^{3}_{2})
= E_{0(1)} + E_{0(2)}, \eqno(5.6)$$
such that $\Delta E_{0} = E_{0} - (E_{0(1)} + E_{0(2)}) > 0$,
which allows  creating two solitons. In contrast, in the NLS case,
the sign of the rest energy is negative and instead of inequality (5.6) 
one has $E_{0} < E_{0(1)} + E_{0(2)}$, which forbids
 decay of the bright soliton. 
\par 
\bigskip\par\noindent
{\bf 6. Resonance interaction of solitons.}
\bigskip\par  
Rewriting RD system (2.3)
in the bilinear form
$$(\pm D_{t} - D^{2}_{x})(G^{\pm} \circ F) = 0,\,\,\,\,
D^{2}_{x}(F \circ F) = - 2G^{+} G^{-},\eqno(6.1)$$
where three new real functions are defined by
$e^{\pm} = (-8 /\Lambda)^{1 \over 2}G^{\pm}/F$,
with the product 
$e^{+}e^{-} =  -{8 \over -\Lambda}(\log F)_{xx}$,
we  apply the Hirota bilinear approach to 
construct solutions of (2.3).  
\par
The one-dissipaton is given by following solution of the 
system (6.1)
$$G^{\pm} = \pm e^{\eta^{\pm}_{1}},\,\,F = 1 + e^{\eta^{+}_{1} + 
\eta^{-}_{1} + \phi_{1,1}},
\,\,
e^{\phi_{1,1}} = (k^{+}_{1} + k^{-}_{1})^{-2},\eqno(6.2)$$
where $\eta^{\pm}_{1} \equiv k^{\pm}_{1} x \pm (k^{\pm}_{1})^{2} t
  + \eta^{\pm (0)}_{1}$, and 
in terms of redefined parameters, $k \equiv (k^{+}_{1} + k^{-}_{1})/2$,
$v \equiv -(k^{+}_{1} - k^{-}_{1})$ it has the form (2.4). 
\par Now we compare  the one-dissipaton boundary conditions with 
the horizon condition (4.5).  
In the space of 
    parameters $(v, k)$ there exist  the
critical value  $v_{crit} =
 2k$. For  solution (6.2) when
$v < v_{crit}$,
   one has 
$e^{\pm} \rightarrow 0$ at infinities. 
So the vanishing b.c. for dissipaton are equivalent to the black hole
(BH) existence.
The corresponding heavy particle we call the BH dissipaton. 
 At the critical value
 the solution is  a kink steady state  in
the moving frame
 $e^{\pm} = \pm k e^{\pm k \xi_{0}}
(1\, \mp \, \tanh \;k\xi )$, with constant
asymptotics
$e^{\pm} \rightarrow \pm \, 2 k e^{\pm k
\xi_{0}}$ for $x \rightarrow \mp \infty$ and
$e^{\pm}
\rightarrow \pm \, 0$ for  $x
\rightarrow \pm \infty$. 
In this case we have the $extremal$ black hole or the 
EBH dissipaton.
In the  over-critical
case $v > v_{crit}$,    $e^{\pm}
\rightarrow \pm \infty$ for $x
\rightarrow \mp \infty$   and  $e^{\pm}
\rightarrow \pm \, 0$ for  $x
\rightarrow \pm \infty$, no black hole exists
and we have very fast and light particles called the  LD. 
\par
For the two-dissipaton solution we have
$$G^{\pm} = \pm [e^{\eta^{\pm}_{1}} + e^{\eta^{\pm}_{2}}
+ ({\breve k^{\pm\pm}_{12}\over k^{\pm\mp}_{21}k^{+-}_{11}})^{2}
e^{\eta^{+}_{1} + \eta^{-}_{1} + \eta^{\pm}_{2}} +
 ({\breve k^{\pm\pm}_{12}\over k^{\pm\mp}_{12}k^{+-}_{22}})^{2}
e^{\eta^{+}_{2} + \eta^{-}_{2} + \eta^{\pm}_{1}}] ;\eqno(6.3a)$$
\par
$$F = 1 + {e^{\eta^{+}_{1} + \eta^{-}_{1}} \over
(k^{+-}_{11})^{2}}
+ {e^{\eta^{+}_{1} + \eta^{-}_{2}} \over
(k^{+-}_{12})^{2}}
+ {e^{\eta^{+}_{2} + \eta^{-}_{1}} \over
(k^{+-}_{21})^{2}}
+ {e^{\eta^{+}_{2} + \eta^{-}_{2}} \over 
(k^{+-}_{22})^{2}}
$$
$$+ ({\breve k^{++}_{12}\breve k^{--}_{12} \over k^{+-}_{12}
k^{+-}_{21}k^{+-}_{11}k^{+-}_{22}})^{2}
e^{\eta^{+}_{1} + \eta^{-}_{1} + \eta^{+}_{2} + \eta^{-}_{2}},\eqno(6.3b)$$
where
$k^{ab}_{ij} \equiv k^{a}_{i} + k^{b}_{j},\,\,\,\,\,\,
 \breve k^{ab}_{ij} \equiv k^{a}_{i} - k^{b}_{j},$
$\eta^{\pm}_{i} \equiv k^{\pm}_{i} x \pm (k^{\pm}_{i})^{2} t
  + \eta^{\pm (0)}$.
\par
First we consider the degenerate case of (6.3), when
$k^{+}_{1} = k^{-}_{1} \equiv p_{1}$, 
   $k^{+}_{2} = k^{-}_{2} \equiv p_{2}$.
Then the solution can be simplified and after some transformation 
be represented in the following form
$$e^{\pm} = \pm ({8 \over -\Lambda})^{1 \over 2} p_{+}p_{-}
{{p_{1}\cosh \theta_{2} e^{\pm p_{1}^{2}t} + 
p_{2}\cosh \theta_{1} e^{\pm p_{2}^{2}t}}   \over 
{p^{2}_{-} \cosh \theta_{+} + p^{2}_{+} \cosh \theta_{-} 
+ 4p_{1}p_{2} \cosh (p_{+}p_{-} t)}     }, \eqno(6.4)$$
where, $p_{\pm} \equiv p_{1} \pm p_{2}$, $\theta_{\pm} 
\equiv \theta_{1} \pm \theta_{2}$, $\theta_{i} \equiv
p_{i}(x - x _{0i}),\,\, (i = 1,2)$.
It
describes collision of two identical dissipatons with 
amplitudes $p_{+}/2$, moving with
the equal velocities  $|v| = |p_{-}|$ and
creating the resonance state with the 
life time,
$\Delta T \approx 2p_{2} d / p_{+}p_{-}$,
depending of the relative distance $d$, where 
$x_{01} = 0$, $x_{02} = d$. This particular solution was considered 
in Ref. 7.
\par 
In a more general case, when 
 $k^{\pm}_{i} > 0$,
$ (i = 1,2)$, and
$\breve k^{+-}_{11} > 0, \,\, \breve k^{+-}_{22} > 0,\,\,
  \breve k^{+-}_{12} > 0, \,\, \breve k^{+-}_{21} < 0$,
(6.3) describes collision of two different dissipatons with amplitudes
$k^{+-}_{12}/2$ and $k^{+-}_{21}/2$ and velocities 
$v_{12} = -\breve k^{+-}_{12}$ and  $v_{21} = -\breve k^{+-}_{21}$
correspondingly. Depending of the positions shift the resonance states 
can be created. 
\par
As a simplest case we consider conditions for decay of 
a BH dissipaton at the rest ($v = 0$) on two dissipatons (2.4)
with
parameters $(k_{1},v_{1})$ and $(k_{2},v_{2})$. From the mass,
momentum and energy conservation laws we obtain relations
$$v^{2}_{1} = 4 k^{2}_{2},\,\,\, \,v^{2}_{2} = 4 k^{2}_{1}.$$
Two possibilities exist:
\par
a) $|v_{1}| = |v_{2}|$. In this case $|k_{1}| = |k_{2}|$ and 
both dissipatons have the equal mass $M_{1} = M_{2} = M/2$
and velocities, satisfying the extremal conditions  $v^{2}_{i} = 
4 k^{2}_{i},\,(i = 1,2)$ and corresponding to  two EBH illustrated
on Fig.1 (for $t > 30$). 
\par
b) $|v_{1}| > |v_{2}|$ (without lose of generality). In this 
case $v^{2}_{1} > 4 k^{2}_{1}$ and $v^{2}_{2} < 4 k^{2}_{2}$,
and we have decay on the one BH dissipaton and the one LD as illustrated
in Fig.1 (for $t < 0 $).
\par
On Fig.2 the interaction of two BH dissipatons by exchange of LD is
shown. A more complicated interaction (Feynman diagram) with 
4 vertices is shown on Fig.3. Detailed description 
of various interactions  simulated by MATHEMATICA would be
published elsewhere. 
\bigskip\par\noindent
{\bf 7. Black holes as topological solitons.} 
\bigskip\par
The gauge equivalence between (2.3) and (3.6) allows us 
to construct exact solutions of (3.6)
providing simple geometrical visualization of the event horizon 
position and to treat
the  black hole as  topological soliton. Under collision they
should have similar resonance properties as solutions of Eqs. (1.2) or (2.3).
In the matrix representation for $SO(2,1)$
$$S = i \left(\matrix{S^{3} & S^{-} \cr
                      S^{+} & -S^{3}\cr}   \right) \,\,=
\,\, ({\bf s}, {\bf \tau})\,\,=\,\, g \tau_{3} g^{-1}, \eqno(7.1)$$
where $S^{\pm} = S^{1} \pm S^{2}$, $S^{2} = - I$, $\det S = 1$,
Eq.(3.6) has the standard matrix form
$$ \partial_{t} S = {1 \over 2i}[S, \partial^{2}_{x} S],\eqno(7.2) $$
with the Lax pair
$$ J^{HM}_{1} = {i \over 4} \lambda S,\,\, 
J^{HM}_{0} = {i \over 8}\lambda^{2} S
+ {\lambda \over 4} S \partial_{x} S. \eqno(7.3)$$ 
This model is gauge equivalent to the resonance NLS (1.2) and RD (2.3). 
Although the Lax pair for the first one can be derived$^{17}$, for the 
second one it has simple Zakharov-Shabat form
$$ J^{RD}_{1} = \left(\matrix{ {1 \over 4}\lambda & q^{-} \cr
                                 q^{+} & -{1 \over 4} \lambda \cr} \right),
\,\, J^{RD}_{0} = \left(\matrix{{1 \over 8}\lambda^{2} - q^{+}q^{-} &
-(\partial_{x} - {1 \over 2}\lambda)q^{-} \cr
(\partial_{x} + {1 \over 2}\lambda)q^{+} & - {1 \over 8}\lambda^{2} + 
q^{+}q^{-}\cr
   } \right),\eqno(7.4)$$
where $q^{\pm} \equiv ({-\Lambda \over 8})^{1 \over 2}e^{\pm}$.
The gauge equivalence is implemented in the standard way 
by non-Abelian transformation
$J^{HM}_{\mu} = g J^{RD}_{\mu} g^{-1} - \partial_{\mu} g\, g^{-1},
\,\,(\mu = 0,1)$
where the matrix $g(x,t)$ is solution of the linear problem
$\partial_{\mu} g = g\, J^{RD}_{\mu}(\lambda = 0)$.
To construct the magnetic analog of dissipaton solution (2.4) we solve this
system first. The result is
$$g(x,t) = \left(\matrix{\tanh z - \gamma & {1 \over \cosh z}e^{\gamma z 
- k^{2}(1 - \gamma^{2})t} \cr
- {1 \over \cosh z}e^{-\gamma z 
+ k^{2}(1 - \gamma^{2})t} & \tanh z + \gamma \cr
           } \right), \eqno(7.5)$$
where $z \equiv k(x - vt)$, $\gamma \equiv v/2k$,
$\det g = 1 - \gamma^{2}$.
Thus, from (7.1) we find one magnetic dissipaton solution with components
$$S^{3} = -1 + {2 \over 1 - \gamma^{2}}\cosh^{-2} z, \eqno(7.6a)$$
$$S^{-} = {2 \over 1 - \gamma^{2}}\,\cosh^{-1} z \,(\tanh z - \gamma)\,
e^{\gamma z 
- k^{2}(1 - \gamma^{2})t}, \eqno(7.6b)$$ 
$$S^{+} = {2 \over 1 - \gamma^{2}}\,\cosh^{-1} z \,(\tanh z + \gamma)\,
e^{-\gamma z 
+ k^{2}(1 - \gamma^{2})t}. \eqno(7.6c)$$ 
It describes magnetic (curved) analog of dissipaton (2.4).
Indeed, in the moving frame with velocity $v$ the $S^{3}$ component is
time independent while the $S^{-}$ and $S^{+}$ components are decaying 
and growing in time. As well as for dissipatons, properties of solution (7.6)
 essentially depend on
the velocity $v$. We have following cases. 
\par a) $\gamma^{2} < 1$, or $|v| < 2|k|$, 
then $-1 \le S^{3} \le {1 + \gamma^{2}
\over 1 - \gamma^{2}}$.
At any fixed time $t$, $S^{+}$ and
 $S^{-}$ vanish when $z \rightarrow \pm\infty$,
while $S^{3} \rightarrow -1$. Due to ${\bf s} = (0,0,-1)$ 
when $z \rightarrow \pm\infty$,
the real line R is compactified. Since the hyperboloid 
${\bf s}^{2} = - (S^{1})^{2} + (S^{2})^{2}  - (S^{3})^{2} = -1$
has topology of cylinder $R \times S^{1}$, the solution (7.6) describes the 
mapping $S^{1} \rightarrow S^{1}$ with degree one. Thus, we see that (7.6)
is the topological soliton, travelling with constant velocity $v$.
When $z = z_{H}$, so that $\tanh z = \pm \gamma$, one of the components
$S^{+}$ or $S^{-}$ is zero. According to (3.7), in this case the metric
component $g_{00} = 0$ and like dissipatons we have the event horizon
at (4.5)
(see Fig.4).
Since any topological soliton configuration cross each of the lines $S^{+} = 0$
and $S^{-} = 0$ at least once, the intersection points correspond to the
event horizon. This fact relates the black holes solution with topological
soliton on hyperboloid. 
\par
b) $\gamma^{2} > 1$, or $|v| > 2|k|$. 
In this case $-{1 + \gamma^{2}\over 1 - \gamma^{2}}
 \le S^{3} \le -1$. At $z \rightarrow \pm\infty$ one of the components
$S^{+}$ or $S^{-}$ is exponentially growing. Moreover, asymptotics at 
$+\infty$ and $-\infty$ are orthogonal. This is why the solution (7.6) has
trivial winding. In fact it can be considered as the turning travelling wave,
which never cross the asymptotic lines at finite distance. Thus, in this 
case there is no event horizon and black holes. 
\par From the above analysis we can see that existence of black holes and 
event horizon intimately related to the topologically nontrivial
solutions of the model (3.6).
\bigskip
{\bf 8. Conclusions}
\bigskip\par
The black hole picture with resonance interaction described above can 
be applied to physical models of slowly varying quasimonochromatic wave
in nonlinear media with the sign indefinite dispersion. 
The  1+1 dimensional case
can be realized particularly in the nonlinear optics.  
The great variety of optical solitons is due to many different properties
of the media involved, including nonlinearity, material and geometric dispersion,
passive or active properties, etc.  
The crucial role in soliton properties play the group-velocity
dispersion of the optical fibers, which depends  not only on the
property of glass material, but also on the waveguide property 
of the fiber. 
According to the sign of dispersion (positive or negative(anomalous)),
two types of NLS are known - defocusing and focusing cases, 
which admit the ''dark" and the ''bright" solitons correspondingly.
Since the envelope wave function 
is a complex quantity, the quadratic dispersion
in general consists of two parts: the phase dispersion 
and the modulus dispersion. The first 
one corresponds to the geometrical optics, while the second one is
responsible for  the diffraction. But in both focusing and defocusing cases,
contribution to dispersion from the phase and the  modulus has the same sign
 (positive
and negative correspondingly).  
In the present paper we considered the nonlinear media
with the sign indefinite quadratic dispersion, which is the result 
of competition between contributing 
with the opposite signs of
the phase and the modulus dispersions.  Not going into microscopic details 
we confined ourself here only by phenomenological description of
some  hypothetical medium. 
Then, the response of the medium to an action of a
quasimonochromatic wave with complex amplitude $\psi(x,t)$, which is slowly 
varying function of the coordinate and the time, is described by 
our resonance version of NLS. 
If it can be realized experimentally, 
the resonance collisions of solitons would be an
interesting tool in optical communication systems with the black hole physics 
flavor. 
\bigskip\par\noindent
{\bf Acknowledgments}
\par
The authors would like to thank Prof. M. Kruskal for useful discussions.
This work was supported in part by the NSC at Taipei and Institute of
Mathematics, AS, Taiwan.   
\bigskip\par\noindent
{\bf References}   
\bigskip  
\bigskip\par
\par
1. W. G. Unruh, {\it Phys. Rev. Lett.}
{\bf 46}, 1351 (1981); {\it Phys. Rev. D }{\bf 51}, 2827 (1995).
\par
2. M. Visser, {\it Class. Quant. Grav.} {\bf 15}, 1767 (1998).
\par
3. T. A. Jacobson and G. E. Volovik, ``Event horizons and ergoregions in $^3$He",
cond-mat/9801308.
\par 
4. R. Jackiw, in {\it Quantum Theory of Gravity}, S. Christensen, 
ed. (Adam Hilger,
Bristol, 1984), p.403; C. Teitelboim, {\it ibid}, p. 327.
\par
5. J. Gegenberg and G. Kunstatter, ``The geometrodynamics of Sine-Gordon 
solitons", hep-th/9807042.
\par 
6. L. Martina, O. K. Pashaev and G. Soliani, {\it Class. Quant. Grav.} {\bf 14},
3179 (1997); O. K. Pashaev, {\it Nucl. Phys.} B (Proc. Suppl.) {\bf 57}, 338
(1997).
\par
7. L. Martina, O. K. Pashaev and G. Soliani, {\it Phys. Rev. D} 
{\bf 58}, 084025
(1998).
\par
8. L. D. Landau and E. M. Lifshitz, {\it Statistical Physics, Part 2},
(Pergamon Press, Oxford, 1980).
\par
9. E. Madelung, {\it Z. Phys.} {\bf 40}, 332 (1926).
\par
10. L. de Broglie, {\it C.R. Acad. Sci. }(Paris), {\bf 183}, 447 (1926).
\par 
11. D. Bohm,{\it Phys. Rev. }{\bf 85}, 166 (1952).
\par
12. E. Nelson, {\it Phys. Rev.} {\bf 150}, 1079 (1966).
\par
\par
13. G. Salesi, {\it Mod. Phys. Lett. }{\bf A22}, 1815 (1996);
G. Salesi and E. Recami, ``Hydrodynamics of spinning particles",
hep-th/9802106.
\par
14. S. A. Akhmanov, A. P. Sukhorukov and R. V. Khokhlov,
{\it Soviet Physics Uspekhi}, {\bf 93}, 609 (1968).
\par
15. M. Cadoni, ''2D Extremal Black Holes as Solitons", hep-th/9803257.
\par
16. C. Callan, S. Giddings, A. Harvey and A. Strominger, {\it Phys. Rev. D},
{\bf 45}, 1005 (1992).
\par
17. O. K. Pashaev, J. -H. Lee and C.-K. Lin, ``Dissipative and Envelope Solitons
Equivalence relations", (to be published).
\vskip1.3cm
\vfil
\centerline{\bf Figures}
\vfil
\vskip1.25cm
\vfil
\item{\bf Fig. 1a.}
{3D plot of two dissipatons resonance-type collision for $k^{+}_{1} = 0.1$, 
$k^{-}_{1} = 1$, $k^{+}_{2} = 1$, $k^{-}_{2} = 0$ and $d = 30$ in the (x,t) plane.}
\vskip1.25cm
\vfil
\item{\bf Fig. 1b.}
{Contour plot of two dissipatons collision with BH resonance in the (x,t) plane.}
\vskip1.3cm
\vfil
\item{\bf Fig. 2a.}
{3D plot of two BH dissipatons exchange-type collision for $k^{+}_{1} = 2$, 
$k^{-}_{1} = 1$, $k^{+}_{2} = -1.7$, $k^{-}_{2} = -1.9$ and $d = 50$.}
\vskip1.25cm
\vfil
\item{\bf Fig. 2b.}
{Contour plot of two BH dissipatons exchange-type collision 
in the (x,t) plane.}
\vskip1.25cm
\vfil
\item{\bf Fig. 3.}
{Contour plot of two-BH dissipatons 4 vertex-type collision  for $k^{+}_{1} = 2$, 
$k^{-}_{1} = 1$, $k^{+}_{2} = 1$, $k^{-}_{2} = 0.3$ and $d = 30$ in the (x,t) plane.}
\vskip1.25cm
\vfil
\item{\bf Fig. 4.}
{Parametric plot of topological soliton projection on 
$(S^{1},S^{2})$ plane for $k = 1$, 
$\gamma = 0.5$. 
Positions of the BH 
event horizon correspond to intersection points with $S^{+} =0$
and $S^{-} = 0$ lines, while asymtotics on $\pm \infty$ to the begining of 
coordinates $(0,0)$.  }
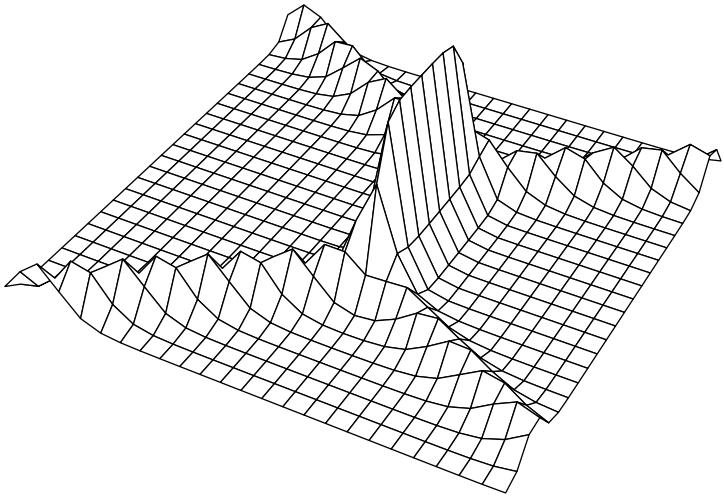
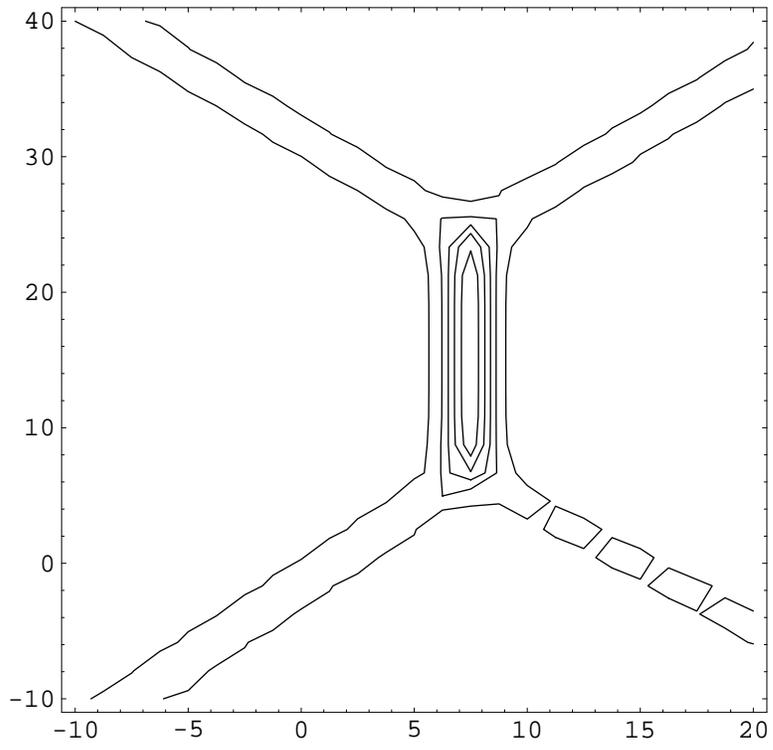
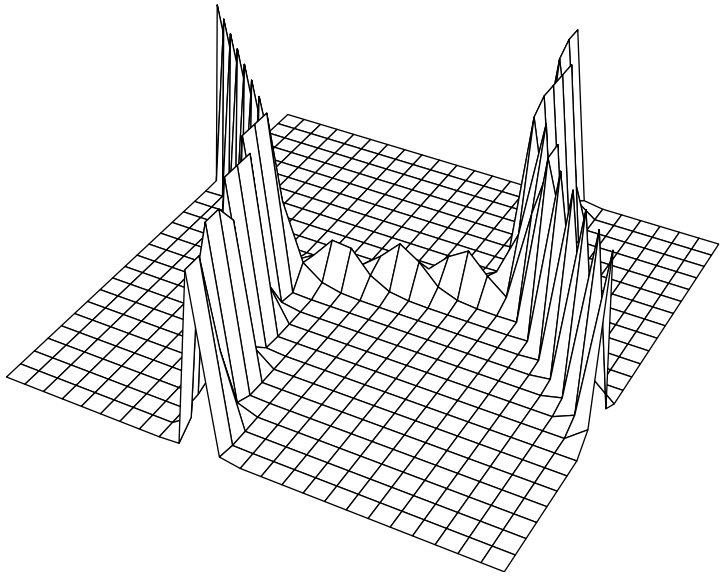
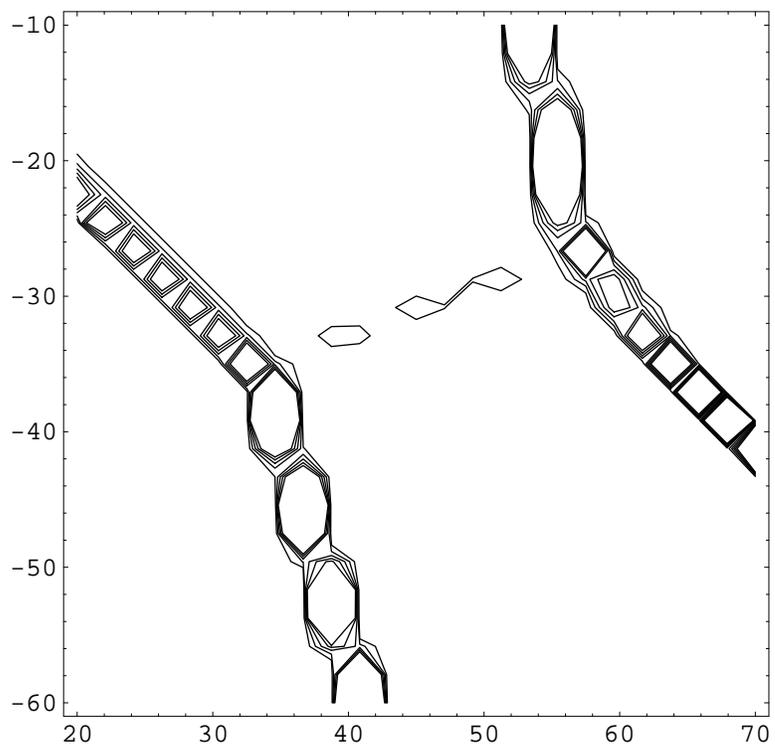
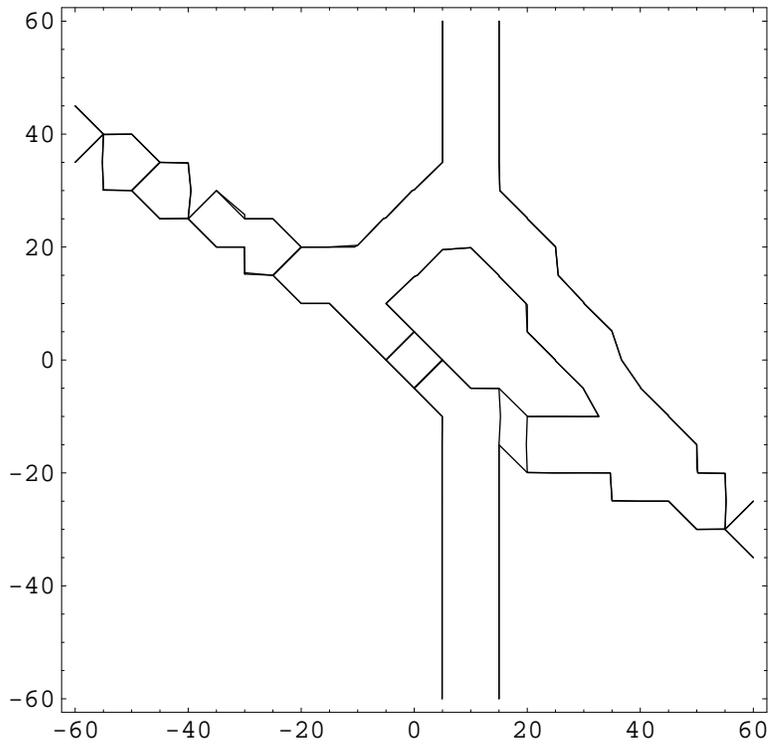
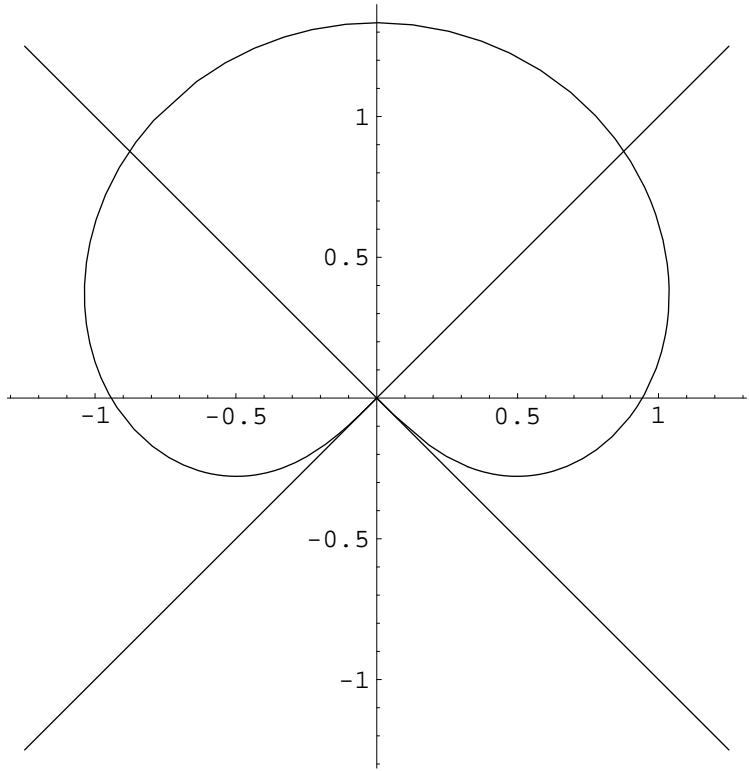
\end